\documentclass[twocolumn]{aastex631}
\usepackage{amsmath}

\newcommand{\hcop}{HCO$^+$}
\newcommand{\kms}{km\,s$^{-1}$}
\newcommand{\ps}{s$^{-1}$}
\newcommand{\naoj}{National Astronomical Observatory of Japan, 2-21-1 Osawa, Mitaka, Tokyo 181-8588, Japan}
\newcommand{\sokendai}{Astronomical Science Program, Graduate Institute for Advanced Studies, SOKENDAI, 2-21-1 Osawa, Mitaka, Tokyo 181-1855, Japan}
\newcommand{\utastro}{Department of Astronomy, The University of Tokyo, 7-3-1 Hongo, Bunkyo-ku, Tokyo 113-0033, Japan}
\newcommand{\asiaa}{Institute of Astronomy and Astrophysics, Academia Sinica, 11F of AS/NTU
Astronomy-Mathematics Building, No.1, Sec. 4, Roosevelt Rd, Taipei 10617, Taiwan}
\newcommand{\shibaura}{Materials Science and Engineering, College of Engineering, Shibaura Institute of Technology, 3-7-5 Toyosu, Koto-ku, Tokyo 135-8548, Japan}
\graphicspath{{./}{figures/}}

\begin{document}

\title{A temperature or FUV tracer? The HNC/HCN ratio in M83 on the GMC scale}

\author[0000-0002-6824-6627]{Nanase Harada}
\affiliation{\naoj}
\affiliation{\sokendai}

\author[0000-0002-2501-9328]{Toshiki Saito}
\affiliation{\naoj}

\author[0000-0003-0563-067X]{Yuri Nishimura}
\affiliation{\utastro}

\author[0000-0002-9668-3592]{Yoshimasa Watanabe}
\affiliation{\shibaura}

\author[0000-0001-5187-2288]{Kazushi Sakamoto}
\affiliation{\asiaa}


\begin{abstract}
The HNC/HCN ratio is observationally known as a thermometer in Galactic interstellar molecular clouds. A recent study has alternatively suggested that the HNC/HCN ratio is affected by the ultraviolet (UV) field, not by the temperature.    
We aim to study this ratio on the scale of giant molecular clouds in the barred spiral galaxy M83 towards the southwestern bar end and the central region from ALMA observations, and if possible, distinguish the above scenarios. 
We compare the high (40-50\,pc) resolution HNC/HCN ratios with the star formation rate from the 3-mm continuum intensity and the molecular mass inferred from the HCN intensities. 
Our results show that the HNC/HCN ratios do not vary with the star formation rates, star formation efficiencies, or column densities in the bar-end region. In the central region, the HNC/HCN ratios become higher with higher star formation rates, which tend to cause higher temperatures. This result is not consistent with the previously proposed scenario in which the HNC/HCN ratio decreases with increasing temperature. Spectral shapes suggest that this trend may be due to optically thick HCN and optically thin HNC.
In addition, we compare the large-scale ($\sim 200$ pc) correlation between the dust temperature from the FIR ratio and the HNC/HCN ratio for the southwestern bar-end region. The HNC/HCN ratio is lower when the dust temperatures are higher. We suggest from the above results that the HNC/HCN ratio depends on the UV radiation field that affects the interstellar medium on the $\sim100\,$pc scale where the column densities are low. 
\end{abstract}

\keywords{}


\section{Introduction} \label{sec:intro}
Molecular emission is a powerful tool for diagnosing physical conditions in star-forming regions because there are chemical formation/destruction reactions and excitation mechanisms that are associated with these physical conditions. Some abundances or line ratios have been proposed as particular tracers of certain physical parameters such as temperature, density, UV field strength, and cosmic-ray ionization rate. 

The HNC/HCN line or abundance ratio is one such example. \citet{1992A&A...256..595S} have suggested from their observations towards OMC-1 and Orion-KL that the HNC/HCN abundance ratios decrease with the increasing temperature, probably due to reactions whose rates become much faster at high temperatures. The HNC/HCN ratios in dark cold cores are much higher than those in OMC-1, providing further evidence for the temperature dependence \citep{1998ApJ...503..717H}. More recently, \citet{2020A&A...635A...4H} compared the temperature obtained from ammonia transitions with the HNC/HCN($J=$1--0) intensity ratios towards the region including OMC-1 to OMC-4, and showed that these two quantities indeed have good correlations. While the HNC/HCN ratio seemed to be a well-established temperature tracer, \citet{2023A&A...679A...4S} recently suggested that it is the strength of the UV radiation field rather than temperature that changes this ratio. High temperature regions often have a UV radiation field because the UV photons heat the gas, which also causes the low HNC/HCN ratio in high temperature regions.

Theoretical calculations of reaction rates also pose challenges to the reliable use of the HNC/HCN ratio as a temperature tracer. Destruction reactions of HNC via atomic oxygen or atomic hydrogen have barriers $\gamma$, and the reaction rates are proportional to exp$(-\gamma/T)$, where $T$ is the gas kinetic temperature. These rates can be fast when the temperature is comparable to $\gamma$, while they are extremely slow at low temperatures. It has been suggested that these reactions with high barriers contribute to the temperature dependence of the HNC/HCN abundance ratio. However, there is a discrepancy between observationally inferred values of barriers and theoretical ones. Observational results suggest that the barriers must be at least a factor of several lower than those obtained by quantum calculations \citep{1996A&A...314..688T,2014ApJ...787...74G} to fit the observed ratio.  

Extragalactic observations have also revealed cases where the HNC/HCN ratio does not necessarily correlate with the temperature. Some galactic nuclei of (ultra-)luminous infrared galaxies that are very bright and warm have been found to have the intensity ratios of HNC/HCN$>1$ \citep{2002A&A...381..783A,2007A&A...464..193A}, which the authors attribute to the infrared pumping or effects of X-rays. In a survey of the center of the starburst galaxy NGC~253, \citet{2022ApJ...939..119B} found a high ($\sim 0.2-1$) HNC/HCN abundance ratio despite the high temperature ($T\sim 50-300\,$K) in this region. A comparison with chemical models suggests that a high cosmic-ray ionization rate ($\sim 10^{-12}\,$\ps) can explain this low ratio despite the high-temperature region. \footnote{We note that this cosmic-ray ionization rate is about one order of magnitude higher than the most probable value derived by \citet{2022ApJ...931...89H} and two orders of magnitude higher than the one estimated from $\gamma$-ray observations \citep{2024MNRAS.tmp..797P} in this region. We caution the readers to note that the cosmic-ray ionization rate can be easily overestimated when the density is overestimated.}

Can the HNC/HCN ratio be used as a temperature tracer? If so, is it still valid for extragalactic sources in different regions of galaxies (center, disk, etc.)? It is helpful to have such a temperature tracer in extragalactic sources because it can be observationally expensive to conduct a large-velocity gradient (LVG) analysis that need multiple transitions. Molecular transitions that provide more accurate temperature values with an LVG analysis (e.g., H$_2$CO and NH$_3$) are also usually much weaker.

To answer the above questions, in this paper, we analyze the HNC/HCN ratio in the center and the bar-end regions of the nearby face-on galaxy M83 using the high spatial resolution ($\sim 50\,$pc) data obtained from ALMA (Atacama Large Millimeter/submillimeter Array). Even HCN observations including the disk region of a galaxy with this spatial resolution are rare beyond the Local Group \citep{2022A&A...660A..83S}. High angular-resolution observations of HNC in the disk region are even rarer. Detection of HNC, which usually has a weaker intensity, is possible in galactic center regions with reasonable observing time in the (sub-)100-pc scale \citep{2005ApJ...618..259M,2019ApJ...884..100H,2022A&A...659A.173E}. With lower spatial resolution, the HNC/HCN ratio in a luminous infrared galaxy has been presented \citep{2018ApJ...855...49H}. However, HNC observations in the disk region with an equivalent spatial resolution are scarce. There has been a spectral scan survey of NGC~1068 that also covers the disk region, which included HNC and HCN, but did not analyze the HNC/HCN ratio \citep{2023ApJ...955...27N}. Our study will provide the most comprehensive view of the HNC/HCN ratio in an extragalactic source to date. 

M83 (also known as The Southern Pinwheel or NGC~5236) is one of the closest face-on barred-spiral galaxies at the distance of $d=4.5\,$Mpc \citep{2003ApJ...590..256T}. It has a moderate total star formation rate of $5.2\,M_{\odot}\,$yr$^{-1}$ \citep{2019ApJS..245...25J}, a factor of a few higher than that of the Milky Way \citep[1.9\,$M_{\odot}$\,yr$^{-1}$;][]{2011AJ....142..197C}. When it comes to the central region, the star formation rate in M83 is about an order of magnitude higher than the Galactic center \citep{2021MNRAS.505.4310C}. The molecular gas mass to fuel the star formation is $2.6\times 10^9\,M_{\odot}$, 15\% of which is concentrated in the central 1\,kpc of this galaxy \citep{2023ApJ...949..108K}. Because of its proximity and its face-on configuration, M83 has been the target of studies at various wavelengths, that revealed various phases of star formation. They include observations of young star clusters from H$\alpha$ images \citep{2011ApJ...729...78W}, supernova remnants \citep{2022ApJ...929..144L},  and atomic gas \citep{2010ApJ...720L..31B,2023A&A...675A..37E}. Abundant molecular gas and the wealth of previous studies and auxiliary data allow for more detailed chemical analysis and comparisons with physical conditions.

This paper is organized as follows. Section \ref{sec:obs} describes the observations and data reduction methods, and we present velocity-integrated images, intensity ratios, selected spectra, and comparison with far-infrared data in Section \ref{sec:results}. These results are discussed in Section \ref{sec:disc}. All of the above is summarized in Section \ref{sec:summary}.

\section{Observations and data reduction} \label{sec:obs}
\begin{figure}
\includegraphics[width=0.49\textwidth]{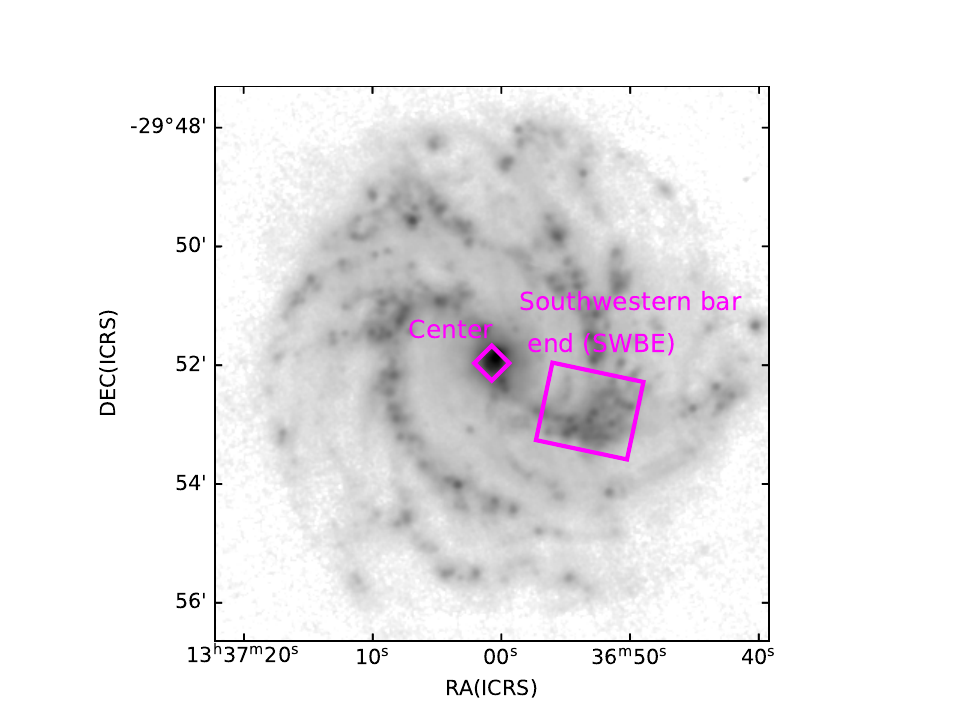}
\caption{Observed regions shown as magenta rectangles superimposed on a Herschel $70\mu$m map of M83 \citep{vngs}. \label{fig:overview}}
\end{figure}

We conduct analyses using $J=1-0$ transitions of HCN, \hcop, and HNC as well as the 3-mm continuum in two different regions of M83, the southwestern bar-end (SWBE) and the central region, using the ALMA data. The observed regions are indicated in Figure \ref{fig:overview}. We also include the \hcop/HCN ratio in our analysis for comparison, as \hcop\ is another commonly observed dense gas tracer and can be simultaneously observed with HCN.

\subsection{M83 southwestern bar end}\label{sec:obsSWBE}
The data towards SWBE are taken as a part of ALMA project code 2021.1.01195.S (PI: Harada), and consist of the 12-meter array data, as well as the Atacama Compact Array (ACA) data (7-meter and total power). This project covers the field of view of $94\arcsec \times 80 \arcsec$ ($2.1 \times 1.8\,$kpc) with a position angle of 78 degrees. 
We imaged the interferometer data (12-meter and 7-meter antennae) with the PHANGS pipeline \citep[ver. 2;][]{2021ApJS..255...19L} using CASA \citep{2007ASPC..376..127M, 2022PASP..134k4501C} version 5.6.3.
After the imaging, we used the CASA version 6.6.0. We used the total power data obtained from the ALMA data delivery (QA2), and used the CASA task \texttt{feather} to combine the interferometry data and the total power data.
We convolved all the cubes into the angular resolution of $1\farcs9 \times 1\farcs9$ (=42\,pc $\times$ 42\,pc) to match the largest beam. Achieved single-channel sensitivities are 30 mK (HCN and \hcop) and 13 mK (HNC) over 3.3 \kms. Primary beam correction was applied to all the cubes. 

We also obtained a 3-mm continuum image at 95.8\,GHz. It has an rms of $10\,\mu$Jy beam$^{-1}$ to estimate the star formation rate. Most of the 3-mm continuum likely originates from free-free emission in HII regions and has a direct relationship with the star formation rate. Because ALMA does not provide the total power capability for the continuum, spatial filtering may be an issue if the continuum emission is extended. Although we cannot check the exact missing flux in the continuum data, we can at least compare the HCN data with and without total power. The HCN emission is usually more extended than the continuum emission. The missing flux of HCN is less than 10\,\% even if we take the aperture of 10.7$\arcsec$, which is the beam size used in Section \ref{sec:glob} to compare with lower-resolution far-infrared data.

Although the 3-mm continuum should be dominated by the free-free emission, it may also contain contamination from the dust emission or synchrotron radiation. We checked the continuum emission at 348.5\,GHz to estimate the contribution from the dust emission, which will be presented in our forthcoming paper. The contamination is $\lesssim 10$\,\% for the region used for our analyses in later sections. The synchrotron radiation should be negligible because there is not many synrotron sources such as supernovae in this region \citep{2022ApJ...929..144L}.

\subsection{M83 center}\label{sec:obsCen}
The HNC data are taken from the cube produced for \citet{2019ApJ...884..100H} (project code: 2016.1.00164.S), and include 12-meter and 7-meter arrays with a maximum recoverable scale (MRS) of $74\arcsec$ at 90 GHz. HCN and \hcop\ data are taken from the ALMA archive \citep[project code 2015.1.01177:][]{2021MNRAS.505.4310C}, which contains only the 12-meter data with an MRS of 25$\arcsec$. These MRSs should be enough to cover most of the emission in the central region judging from the structure of CO(1--0) \citep{2023ApJ...949..108K}. Although the CO(1--0) emission is expected to be much more extended than HCN, \hcop, and HNC, most of the emission comes from regions within the size of 25$\arcsec$. The data were imaged using the same way as \citet{2019ApJ...884..100H} using CASA ver. 5.4 and later processed with CASA ver. 6.6.0. We convolved cubes into a common beam of $2.5\arcsec$(=55\,pc), and applied primary beam correction. Single-channel rms noise levels over 10\,\kms\ are 20 mK, 17 mK, and 10 mK for HCN, \hcop, and HNC, respectively. The continuum image has an rms of 29\,$\mu$Jy beam$^{-1}$.

The 3-mm continuum (more precisely, 2.8-mm or 96.9\,GHz) may also suffer from the dust contamination in addition to the free-free emission here. By comparing the 2.8-mm and 1.0-mm continuum presented in \citet{2019ApJ...884..100H}, we estimate the contamination to be less than 10\,\% for most emission, but can be up to 30\,\% for very low-level ($<0.4\,$mJy\,beam$^{-1}$) emission. The star formation rate corresponding to this level of emission is log($\dot{M}_*$[$M_{\odot}$\,Myr$^{-1}$\,pc$^{-2}$])$<0.5$ (see Section \ref{sec:phys}).

\section{Results}\label{sec:results}
\subsection{Velocity-integrated and continuum images}
\subsubsection{SWBE}
We created velocity-integrated images of HCN, HNC, and \hcop\ applying masks, or intensity thresholds, on cubes. We first convolved HCN and \hcop\ images to twice the beam size, then applied the $2\sigma$ cutoff from the convolved HCN image to create velocity-integrated images of HCN and HNC. The $2\sigma$ cutoff from the convolved \hcop\ image is used to create the \hcop\ image. These reference masks for thresholds were chosen because HNC is much fainter than HCN while \hcop\ is equally as bright as HCN. These velocity-integrated images have different noise levels across the maps because of the cutoff we applied. 

\begin{figure*}
\includegraphics[width=0.95\textwidth]{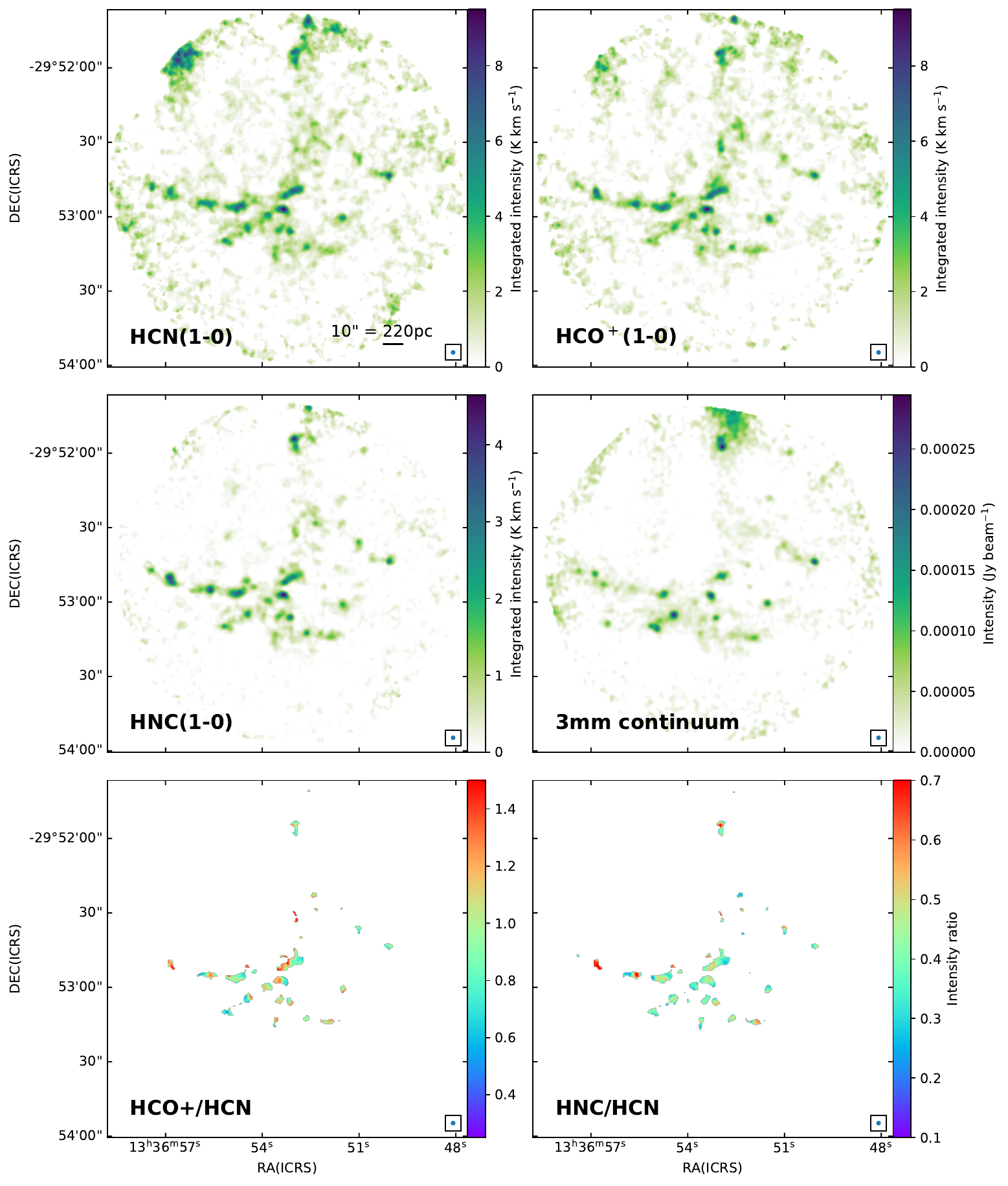}
\caption{Moment 0 images of HCN (upper left), \hcop\ (upper right), and HNC (middle left) and the 3mm-continuum image (middle right) at the southwestern bar end region of M83 and ratio maps of \hcop/HCN (lower left panel) and HNC/HCN (lower right panel). The beam size is shown as a blue circle at the lower right corner of each panel ($1.9\arcsec \times 1.9 \arcsec$ $=42\,\mathrm{pc}\times42\,\mathrm{pc}$). Intensity ratios are derived in the unit of K instead of Jy beam$^{-1}$. \label{fig:mom0_BE}}
\end{figure*}

Figure \ref{fig:mom0_BE} shows the velocity-integrated (moment 0) images of HCN(1--0) in the upper left panel, \hcop(1--0) in the upper right panel, and HNC(1--0) on the middle left panel. The 3-mm continuum is also shown in the middle right panel, which should be dominated by free-free emission from the star formation at this wavelength. These images all show distributions similar to that of CO(1--0) reported by \citet{2023ApJ...949..108K}, tracing the transition from the bar region in the eastern part of the field of view to the spiral arm in the northern part as well as the feather-like structure extending from the bar end to the south or west. The continuum emission shows a slightly more compact and sparse distribution than that of the molecular lines.

\subsubsection{The central region}
\begin{figure*}
\includegraphics[width=0.95\textwidth]{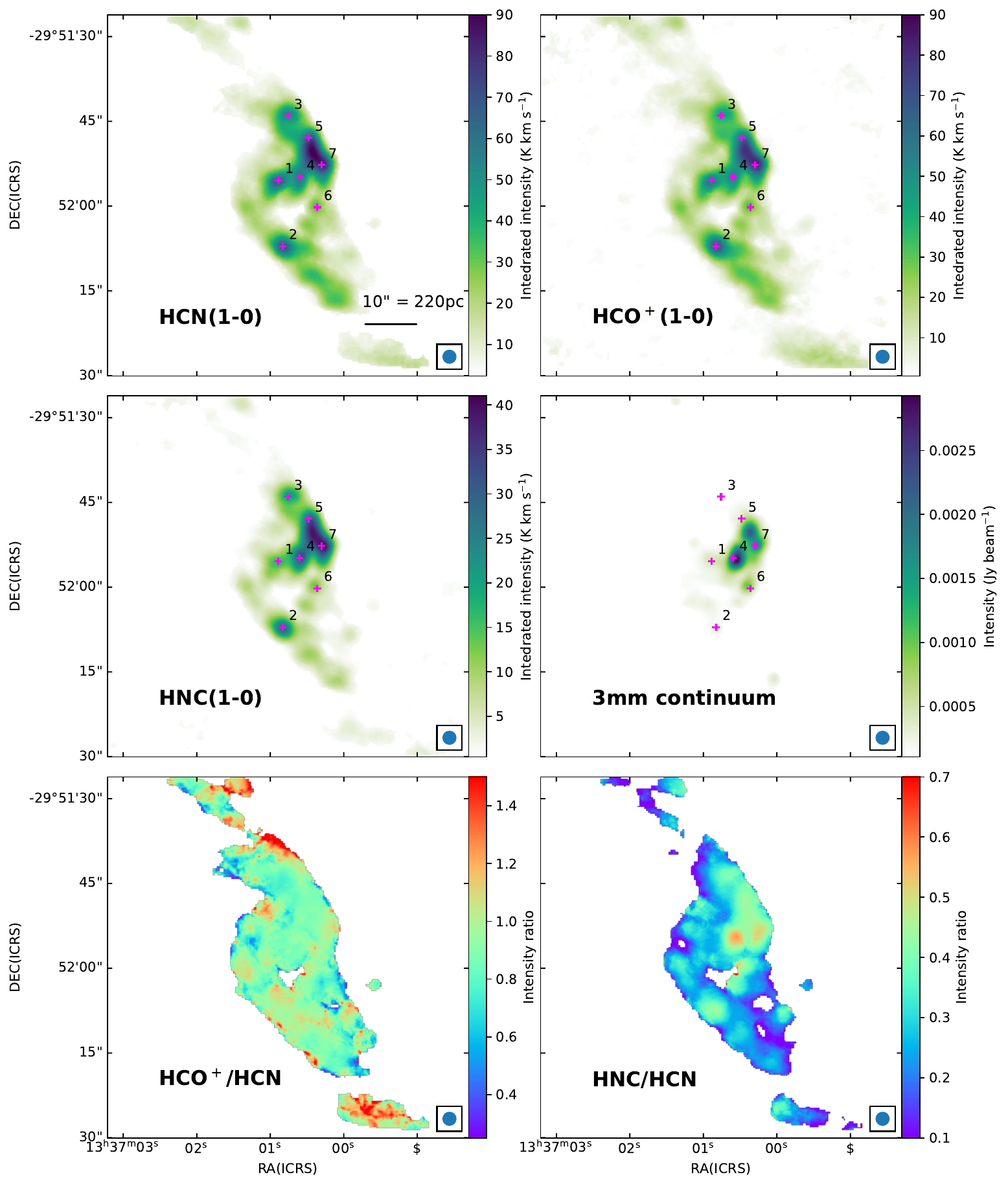}
\caption{Same as Figure \ref{fig:mom0_BE}, but for the central region of M83. The synthesized beam size is $2.5\arcsec \times 2.5 \arcsec$ $=55\,\mathrm{pc}\times55\,\mathrm{pc}$.\label{fig:mom0_cen}}
\end{figure*}

Figure 3 shows images of HCN, \hcop, and HNC integrated intensities and 3 mm continuum emission in the central region of M83. Similar to the images of the SWBE region, we applied the threshold cutoff using the convolved HCN for HCN and HNC, and then convolved \hcop\ for \hcop\ again using $2 \sigma$. The HNC and 3-mm continuum before the beam convolution have already been presented by \citet{2019ApJ...884..100H}, and higher-resolution images of HCN and \hcop\ by \citet{2021MNRAS.505.4310C}. The HCN and \hcop\ images show a remarkable similarity both in low- and high-resolution images. The HNC image shows a similar distribution to that of HCN and \hcop\ as well, but has a weaker emission around the optical nucleus (marked as position 1 in Figure \ref{fig:mom0_cen}). The 3-mm emission is more compact, and appears only in the western part of the central region of M83.

\subsection{Physical conditions of observed regions}\label{sec:phys}
To put the observed region into perspective, we have derived the dense gas surface density  ($\Sigma_\mathrm{dense}$) and the star formation rate surface density ($\Sigma_\mathrm{SFR}$). For the dense gas surface density, we use 
\begin{eqnarray}
\Sigma_{\rm dense} = \alpha_\mathrm{HCN} \int T_b\,dv  \times cos(i) \nonumber \\
= 9.0\,  \mathrm{({\it M}_{\odot}\,pc^{-2}) \int T_b\,dv (K\,km\,s^{-1})^{-1}},
\end{eqnarray}
where $\alpha_{\rm HCN}$ is the conversion factor of $10\, M_{\odot}$\,pc$^{-2}$ (K\,km\,s$^{-1}$)$^{-1}$ \citep{2004ApJS..152...63G}, $\int T_b\,dv$ is the velocity-integrated intensity, and $i$ is the inclination angle \citep[$i=26^{\circ}$;][]{2023ApJ...949..108K}. We derived the star formation rate surface density by first calculating the star formation rate $\dot{M}_*$ within the aperture using the following equation \citep[e.g., ][]{2016MNRAS.463..252B}.
\begin{eqnarray}
\dot{M}_*= 9.49 \times 10^{10} \,M_{\odot} \,{\rm yr}^{-1} g_{\rm ff}^{-1}\left(\frac{\alpha_B}{\rm cm^3\,s^{-1}}\right) \nonumber\\ \times \left(\frac{T_\mathrm{e}}{\mathrm{ K}}\right)^{0.5}  \left(\frac{D}{\rm Mpc}\right) \left(\frac{f}{\rm Jy\,beam^{-1}}\right),
\end{eqnarray}
where $g_{\rm ff}$ is the Gaunt factor, $\alpha_B$ is the effective recombination coefficient, $T_{\rm e}$ is the electron temperature, $D$ is the distance to the galaxy, and $f$ is the flux per beam. We assume the electron temperature to be $T_{\rm e}=5000\,$K. The Gaunt factor is given by 
\begin{equation}
g_{\rm ff} = 0.5535\, {\rm ln} \left[\left(\frac{T_{\rm e}}{\rm \,K}\right)^{1.5}\left(\frac{\nu}{\rm GHz}\right)^{-1}Z^{-1}\right]-1.682,
\end{equation}
where $\nu$ is the frequency and $Z$ is the electric charge, which can be approximated as $\sim 1$.
We approximate the effective recombination coefficient $\alpha_B$ using the following formula \citep{2020ApJ...895...85M}
\begin{equation}
    \alpha_\mathrm{B} = 3.63 \times 10^{-10} \left(\frac{T_{\rm e}}{\rm K}\right)^{-0.79}.
\end{equation}
To derive the star formation rate surface density, we take
\begin{equation}
\Sigma_\mathrm{SFR} (M_{\odot} \,{\rm yr}^{-1} {\rm pc}^{-2})= \dot{M}_*/A \times cos(i),
\end{equation}
where $A$ is the aperture size in pc$^{2}$.

\begin{figure*}
\centering{
\includegraphics[width=0.49\textwidth]{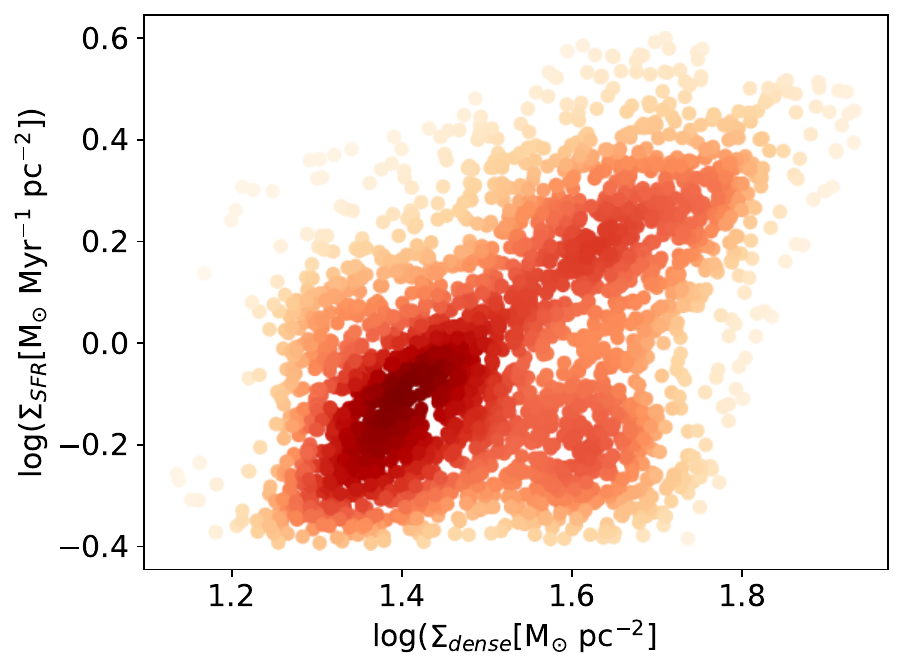}
\includegraphics[width=0.49\textwidth]{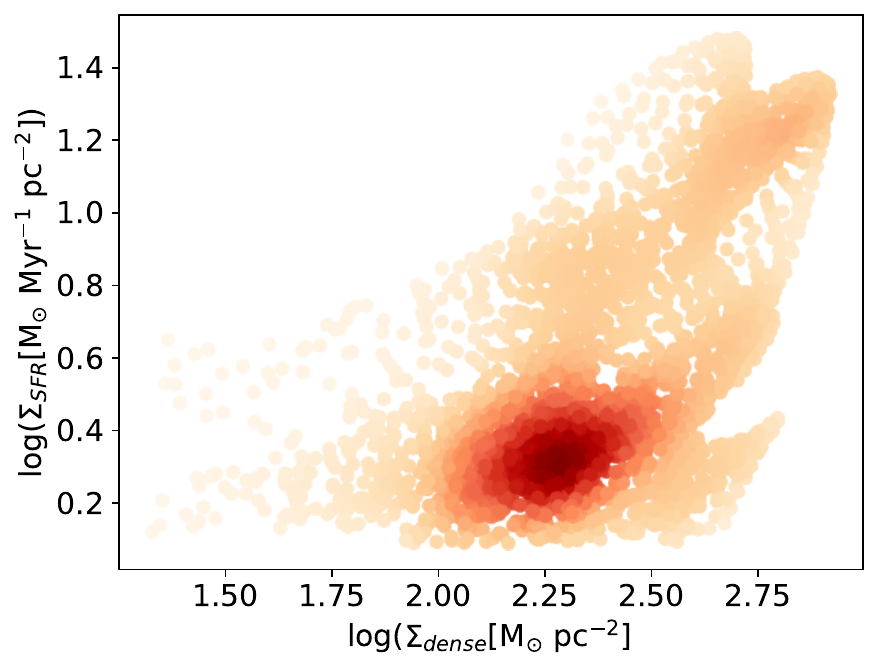}
}
\caption{The star formation rate surface density as a function of the dense gas surface density in the SWBE region (left) and the central region (right). Note that the star formation rate is derived from the 3-mm continuum emission and the dense gas surface density is derived from the HCN intensity. The color of the data point is determined by the density of the data points. Note that, in the right panel, the star formation rate log($\dot{M}_*$[$M_{\odot}$\,Myr$^{-1}$\,pc$^{-2}$])$<0.5$ may contain significant contamination as discussed in Section \ref{sec:obsCen}. \label{fig:SFrelation}}
\end{figure*}

Figure \ref{fig:SFrelation} shows the relationship between the dense gas surface density and the star formation surface density in the SWBE region and the central region. The data points in the SWBE region lie close to or slightly higher than those in the northern arm of M51 \citep{2019A&A...625A..19Q}. Data points from both regions lie roughly in the $\Sigma_{\rm SFR}-\Sigma_{\rm dense}$ relationship in star-forming galaxies \citep{2012A&A...539A...8G,2015AJ....150..115U,2022A&A...660A..83S}. It is fair to say that regions we observe have normal star-forming conditions although the scatter is large due to the high spatial resolution ($\sim 50\,$pc) we used \citep[see the effect of the beam size on the relation in ][]{2022A&A...660A..83S}. As discussed in Section \ref{sec:obs}, the 3-mm continuum has minor contamination from the dust contribution, and caution is needed to interpret the faint 3mm emission (i.e., low star formation rate surface density) in this relation.

\subsection{Intensity ratios}
\subsubsection{SWBE}
The intensity ratio maps, \hcop/HCN and HNC/HCN, of the SWBE region are shown in the lower panels of Figure \ref{fig:mom0_BE}. Only pixels above certain thresholds are shown in the integrated images. The threshold is $7\sigma$ for HCN and \hcop, and $5\sigma$ for HNC. Therefore, the ratios have a significance of at least $5\sigma$ (for \hcop/HCN) and $4\sigma$ (for HNC/HCN). While the \hcop/HCN ratios do not show locations with particularly high or low values (Figure \ref{fig:mom0_BE} lower left), the HNC/HCN ratios have high values ($>0.6$) at the eastern and northern part (Figure \ref{fig:mom0_BE} lower right). 

Figure \ref{fig:ratio_BE} shows the \hcop/HCN (left panels) and HNC/HCN (right panels) ratios as functions of the HCN intensity (top panels), the 3-mm continuum intensity (middle panels), and the 3-mm continuum/HCN ratio (lower panels). The dense gas surface density ($\Sigma_{\rm dense}$), the star formation surface density ($\Sigma_{\rm SFR}$), and the dense-gas star formation efficiency (SFE) corresponding to the quantities used for the x-axes are also shown in the second x-axes. All of the plots show constant \hcop/HCN and HNC/HCN ratios, which seem to have no relations with the HCN intensity, the 3-mm continuum intensity, and the 3-mm/HCN ratio. Overall, the ratios range as follows $I(\mathrm{HCO^+})/I(\mathrm{HCN})=0.99\pm0.33$ and $I({\rm HNC})/I({\rm HCN})=0.42\pm0.12$. Note the \hcop/HCN ratio is higher for the low HCN intensity because the y-axis correlates with an inverse of the HCN intensity, and this is unlikely to be a trend with a physical origin. 

If the HNC/HCN ratio depends on the temperature as suggested by \citet{2020A&A...635A...4H} and other previous works, and if the temperature increases as the star formation rate increases, then the HNC/HCN ratio should decrease as the star formation rate increases. This is not the trend seen in the HNC/HCN ratio vs. the 3-mm continuum intensity relationship (Figure \ref{fig:ratio_BE} middle right). Instead, the HNC/HCN ratio seems to be rather insensitive to the star formation rate.

While the \hcop/HCN ratio appears randomly determined by the noise, there is a cluster of pixels with relatively high HNC/HCN ratios (0.6-0.8) where the HCN or 3-mm continuum has high S/N, corresponding to the eastern and northern parts identified on the ratio map.

\subsubsection{The central region}
\begin{figure*}[tp]
\centering{
\includegraphics[width=0.45\textwidth]{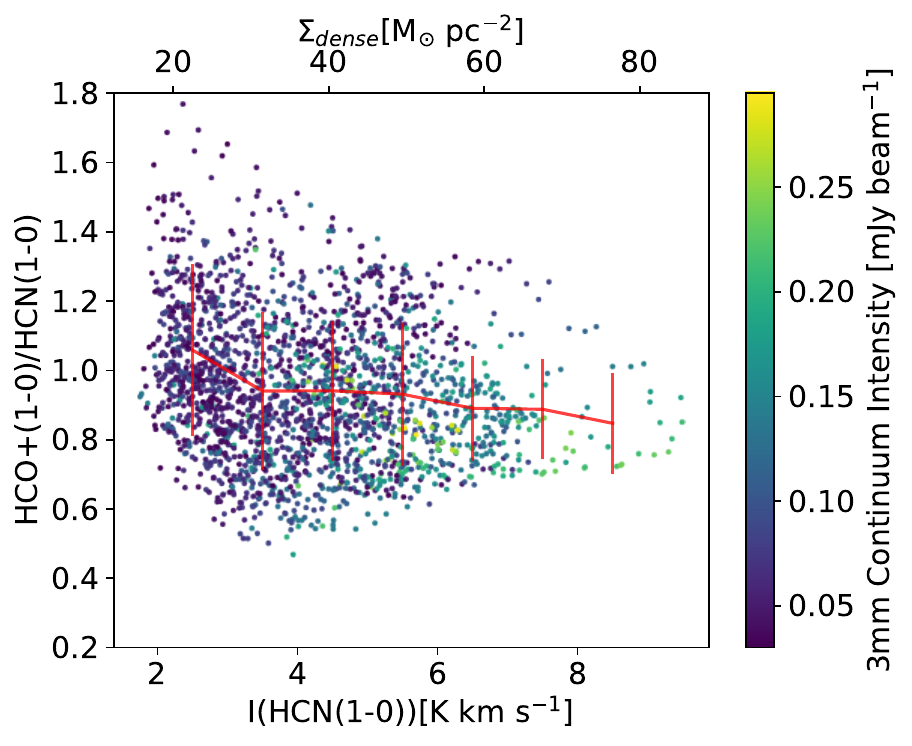}
\includegraphics[width=0.45\textwidth]{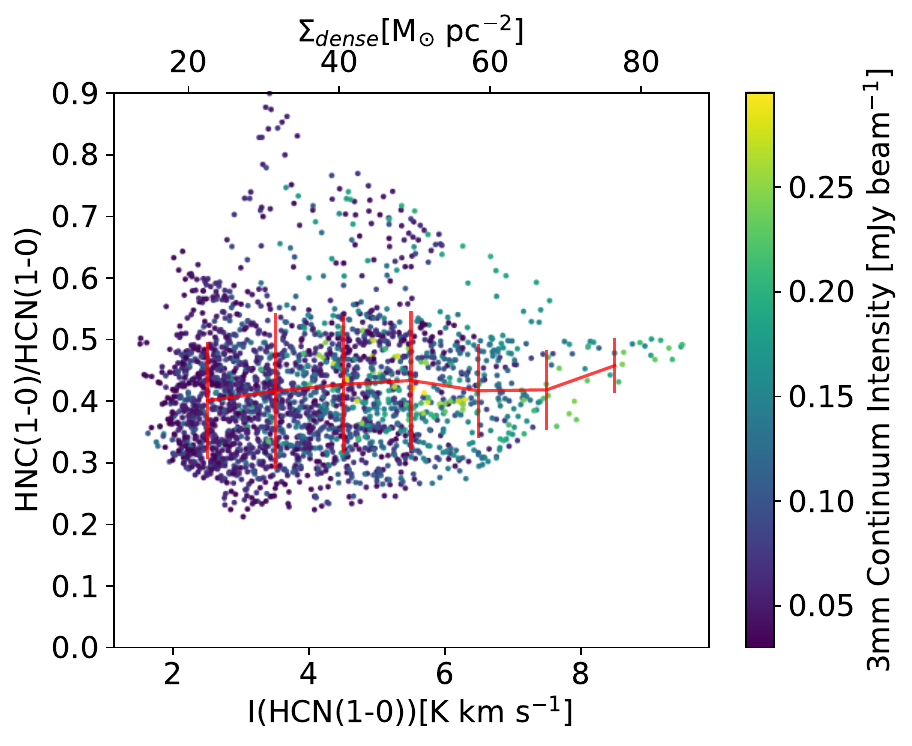}
}
\centering{
\includegraphics[width=0.45\textwidth]{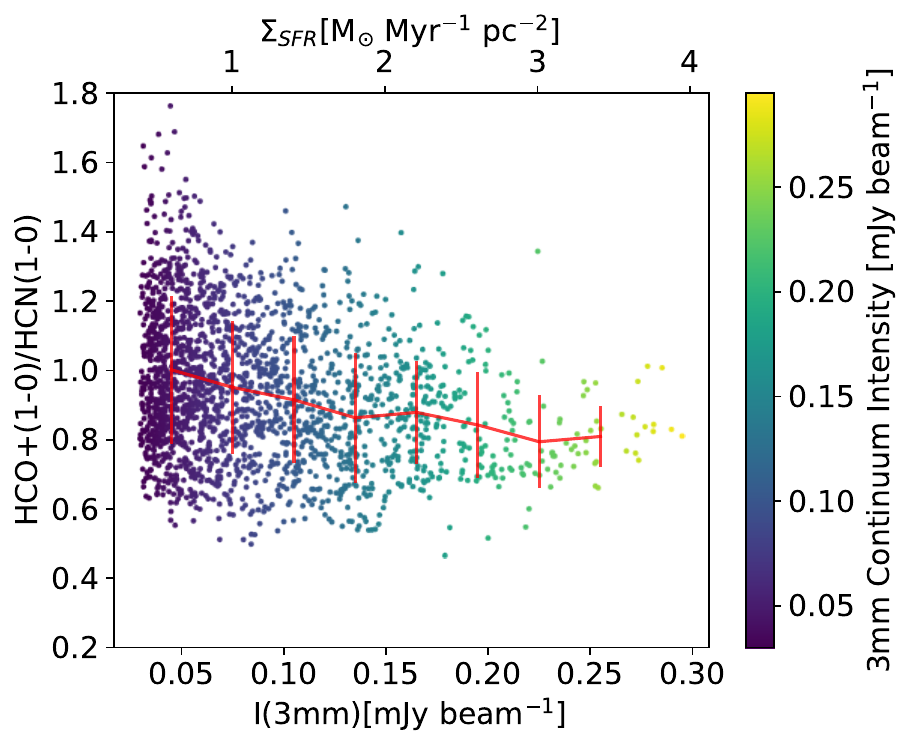}
\includegraphics[width=0.45\textwidth]{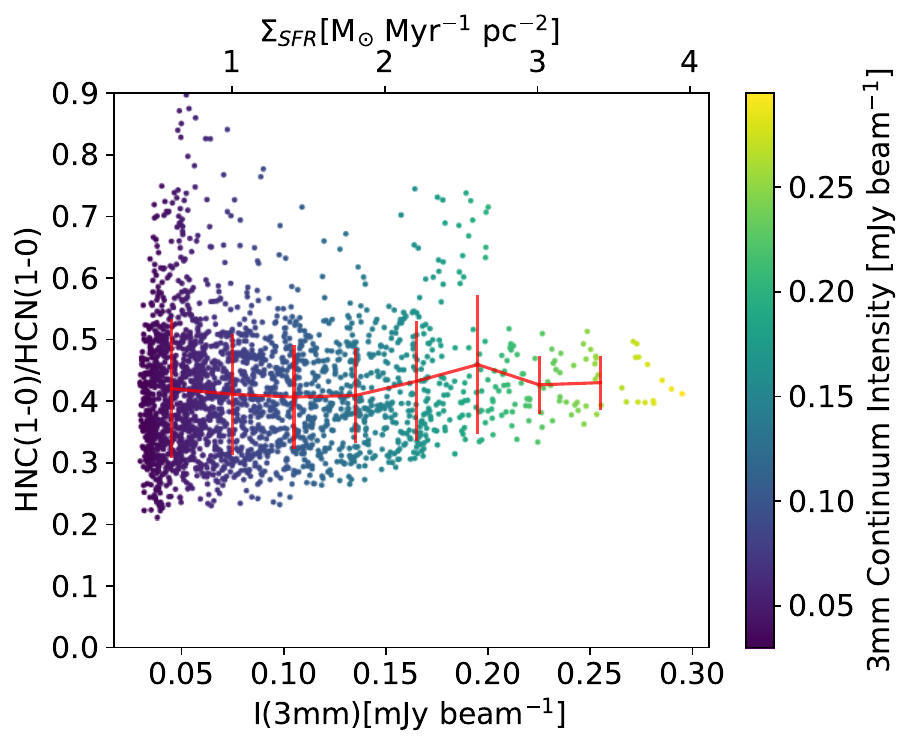}
}
\centering{
\includegraphics[width=0.45\textwidth]{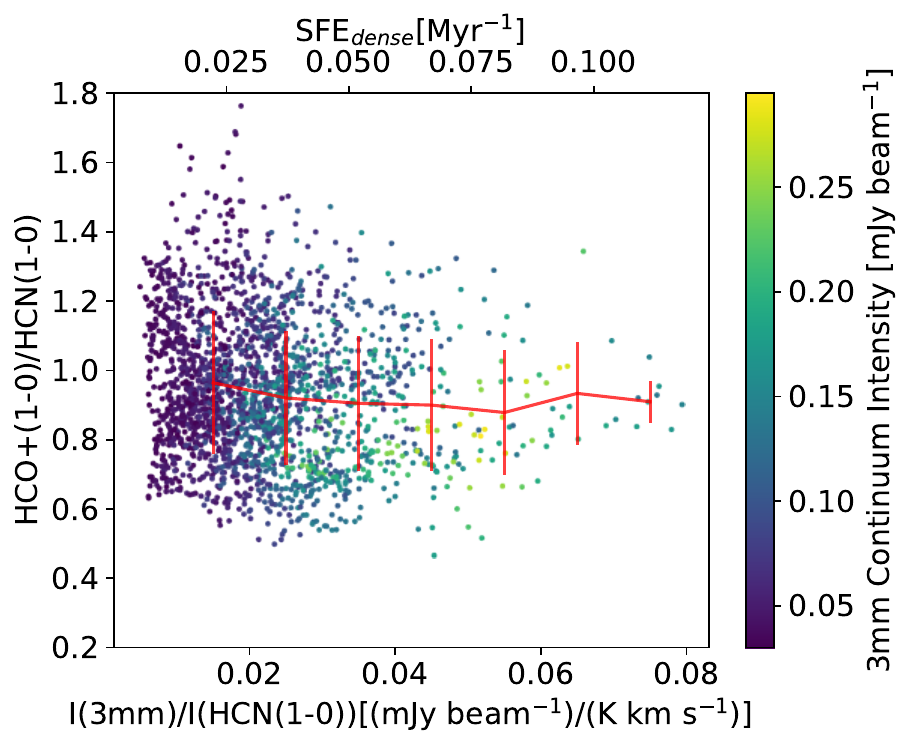}
\includegraphics[width=0.45\textwidth]{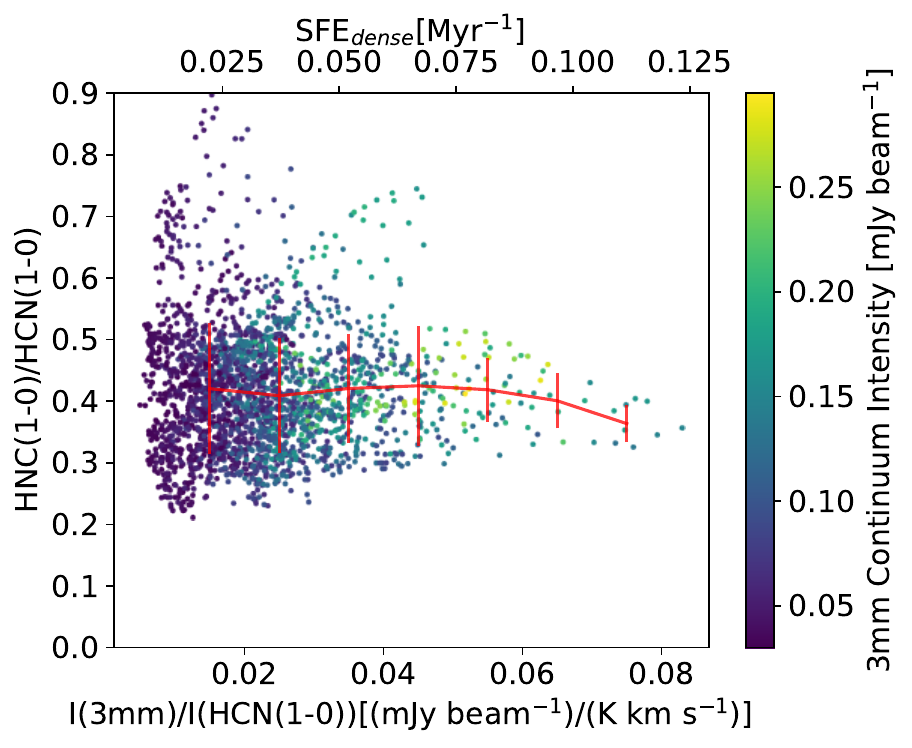}
}
\caption{Intensity ratios of \hcop/HCN(1--0) (left panels) and HNC/HCN(1--0) (right panels) as a function of HCN(1--0) intensity (top panels), 3mm continuum (middle panels), and the intensity ratio of 3mm continuum over HCN(1--0) at the southwestern bar end region of M83. The color represents the 3mm continuum intensity. The standard deviation and the mean of data points in each horizontal bin are shown as red vertical and horizontal lines, respectively. \label{fig:ratio_BE}}
\end{figure*}

\begin{figure*}[tp]
\centering{
\includegraphics[width=0.45\textwidth]{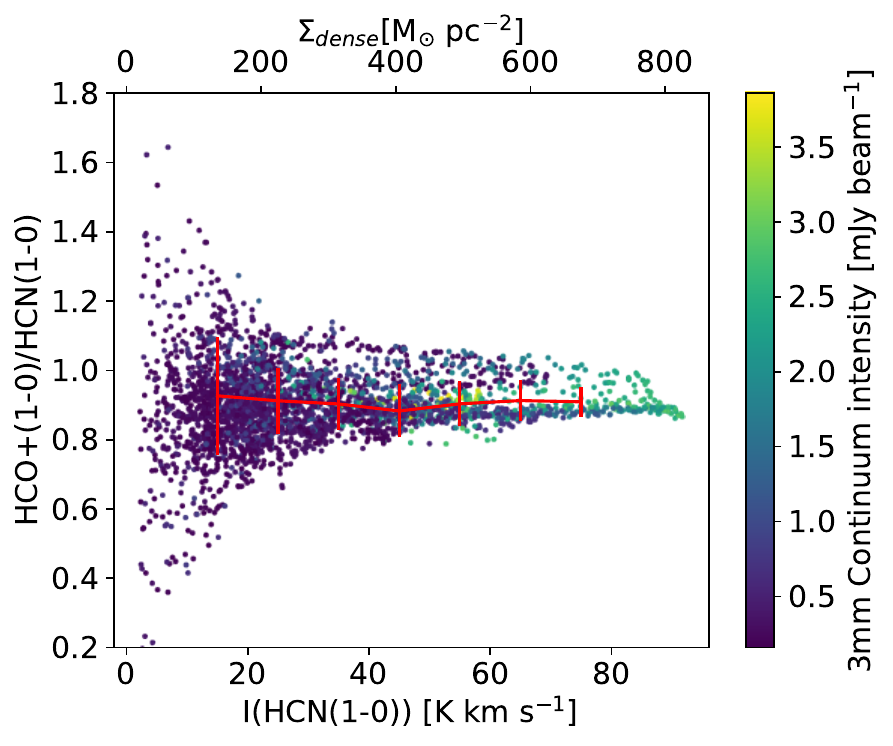}
\includegraphics[width=0.45\textwidth]{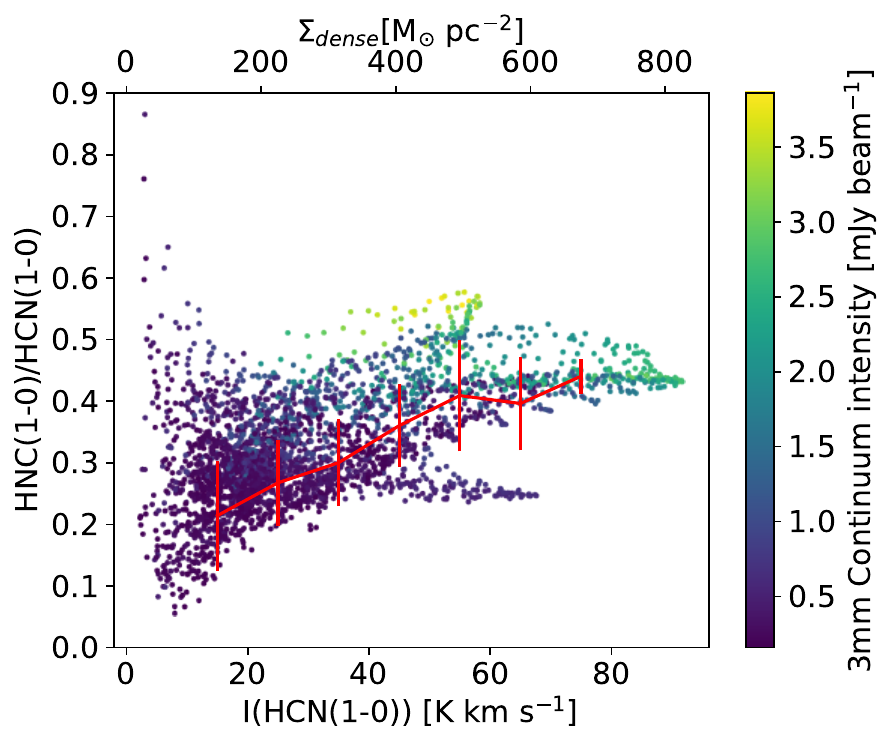}
}
\centering{
\includegraphics[width=0.45\textwidth]{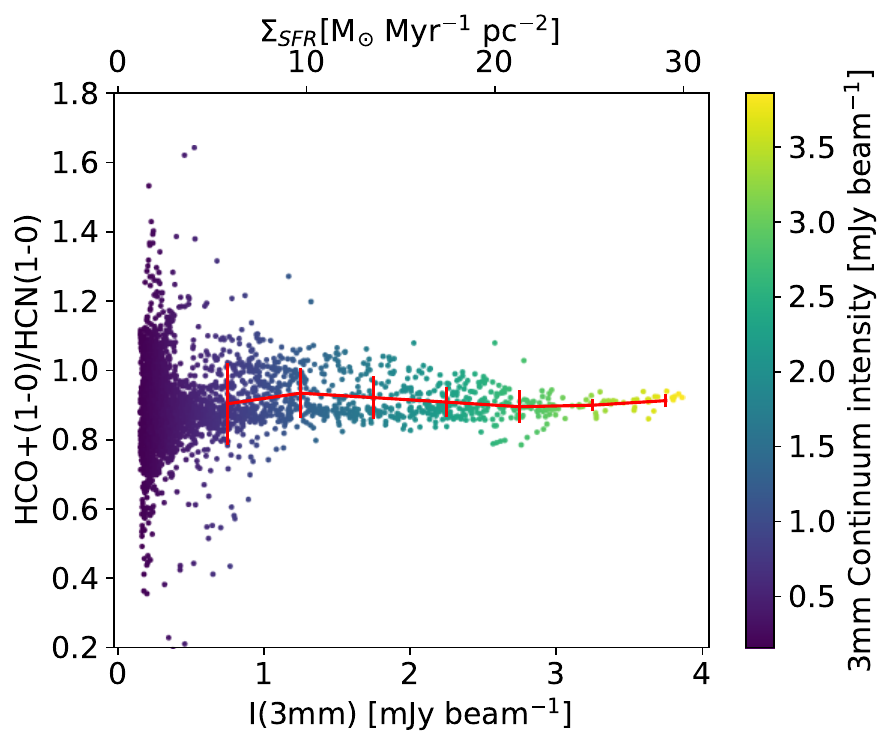}
\includegraphics[width=0.45\textwidth]{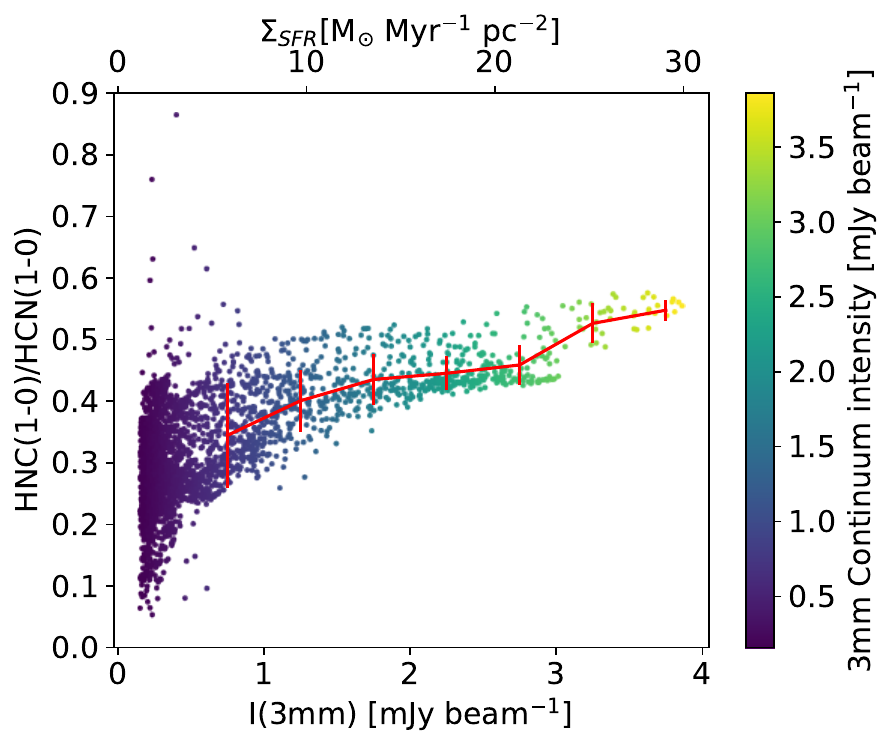}
}
\centering{
\includegraphics[width=0.45\textwidth]{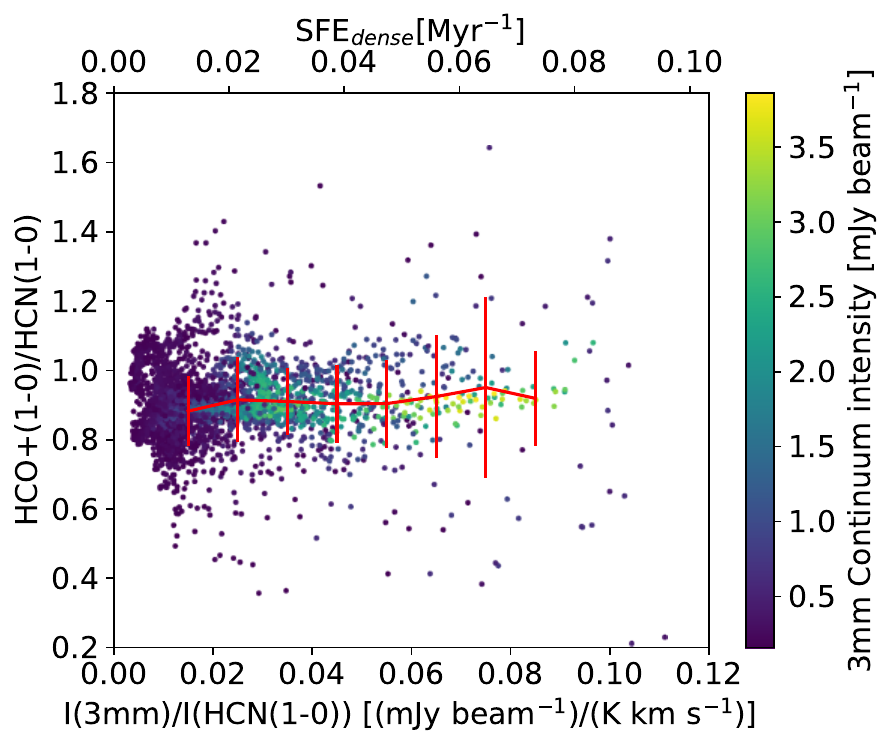}
\includegraphics[width=0.45\textwidth]{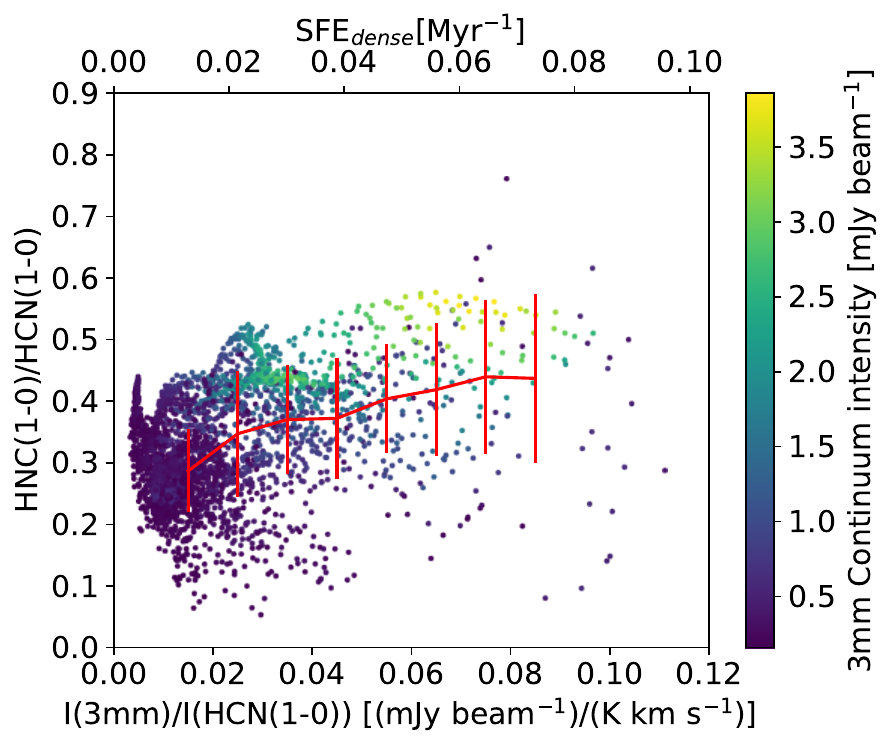}
}
\caption{Same as Figure \ref{fig:ratio_BE}, but for the central region. \label{fig:ratio_cen}}
\end{figure*}

We show ratio maps of \hcop/HCN and HNC/HCN in Figure \ref{fig:mom0_cen} (lower panels). The \hcop/HCN ratio (Figure \ref{fig:mom0_cen} lower left) shows an almost constant ratio except for the deviation caused by the low S/N regions. Meanwhile, the HNC/HCN ratio shows higher ratios in regions where the molecular line emission is brighter. Figure \ref{fig:ratio_cen} shows the same quantities as Figure \ref{fig:ratio_BE} for the central region, and the above trend is confirmed by the relationship between the HNC/HCN ratio and the HCN intensity (Figure \ref{fig:ratio_cen} upper right). The HNC/HCN ratio increases with the increasing HCN intensity. An even clearer trend is seen between the HNC/HCN ratio and the 3-mm continuum intensity (Figure \ref{fig:ratio_cen} middle right). A similar trend is also seen between the HNC/HCN ratio and the 3-mm continuum/HCN ratio (Figure \ref{fig:ratio_cen} lower right). The ratios have means and standard deviations of $I(\mathrm{HCO^+})/I(\mathrm{HCN})=0.92\pm0.21$ and $I({\rm HNC})/I({\rm HCN})=0.26\pm0.11$.

We note that the discrepancy between the ALMA array configurations of the HCN and HNC data is unlikely to be the cause of this trend. If extended emission is missed more in HCN than in HNC due to the lack of the 7-meter array data, then the HNC/HCN ratio would have been lower at low-intensity extended regions. This is the opposite of our results. 

One possible cause of the high HNC/HCN ratio in the high-column-density region is the high optical depth. When there is a large column of molecular gas, the optical depth increases. HCN is more abundant than HNC, and is more likely to be optically thick. To investigate this possibility, we show the spectral shapes of selected positions in Section \ref{sec:spectra}.

\subsection{Spectral shapes}\label{sec:spectra}

\begin{figure*}
\centering{
\includegraphics[width=0.9\textwidth]{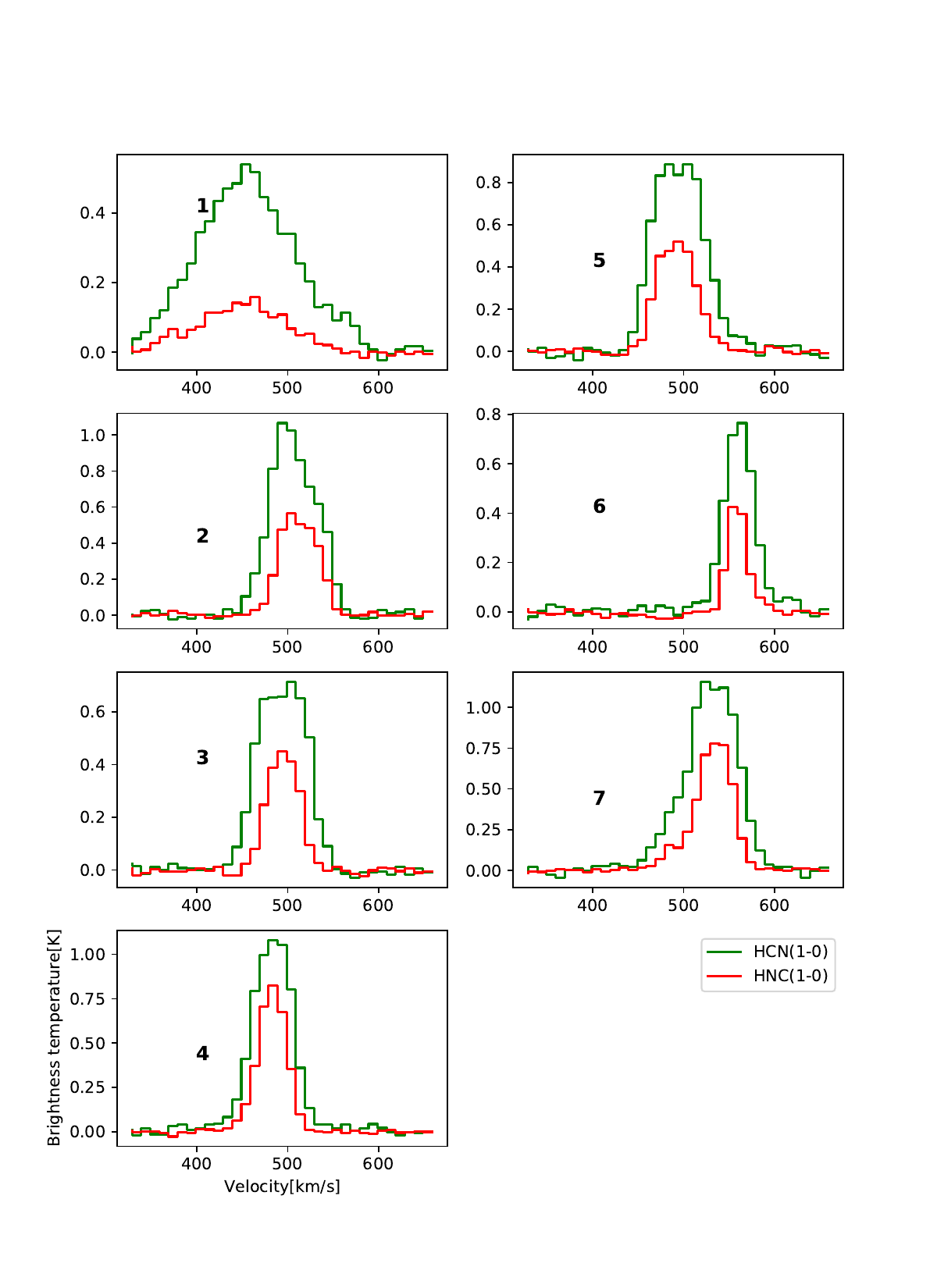}
}
\caption{Spectra of selected regions of HCN (green) and HNC (red) at single points within the 2.5$\arcsec$ beam taken from positions shown in Figure \ref{fig:mom0_cen}. \label{fig:spec}}
\end{figure*}

Figure \ref{fig:spec} shows spectra from positions 1-7 shown in Figure \ref{fig:mom0_cen}. Some spectra show different line shapes between HCN and HNC, specifically at positions 3, 5, and possibly at 7. At these positions, HCN has either flat-top or double-peak profiles while HNC has single-peak Gaussian-like profiles. This suggests that HCN is optically thick compared with HNC. Meanwhile, at position 1, where the HNC/HCN ratio is relatively low, the line profiles of HCN and HNC are similar. This position is close to the optical nucleus \citep{2000A&A...364L..47T}, and their line widths are large. Line emission is likely to be optically thinner at this position than others.

Note that the hyperfine structures of HCN($J=$1--0) alone cannot produce the line profiles we see here. Compared to the strongest transition of $F=2-1$, $F=0-1$ and $F=1-1$ have velocity shifts of $-7.1$\,\kms\ and $4.9$\,\kms, respectively. These hyperfine splittings can cause slightly wider line profiles, but their effect should be small considering the spectral width of our data cube, 10\,\kms. We note that, due to the large line widths (FWHM) in our observations ($\gtrsim 40$\,\kms\ in the central region, $\gtrsim 20$\,\kms\ in SWBE), we cannot check if there is any excitation anomaly among the hyperfine splitting \citep[e.g., ones reported towards Orion-B by ][]{2023A&A...679A...4S}.

\subsection{Global variation of the intensity ratio}\label{sec:glob}
\begin{figure*}
    \centering
    \includegraphics[width=0.99\linewidth]{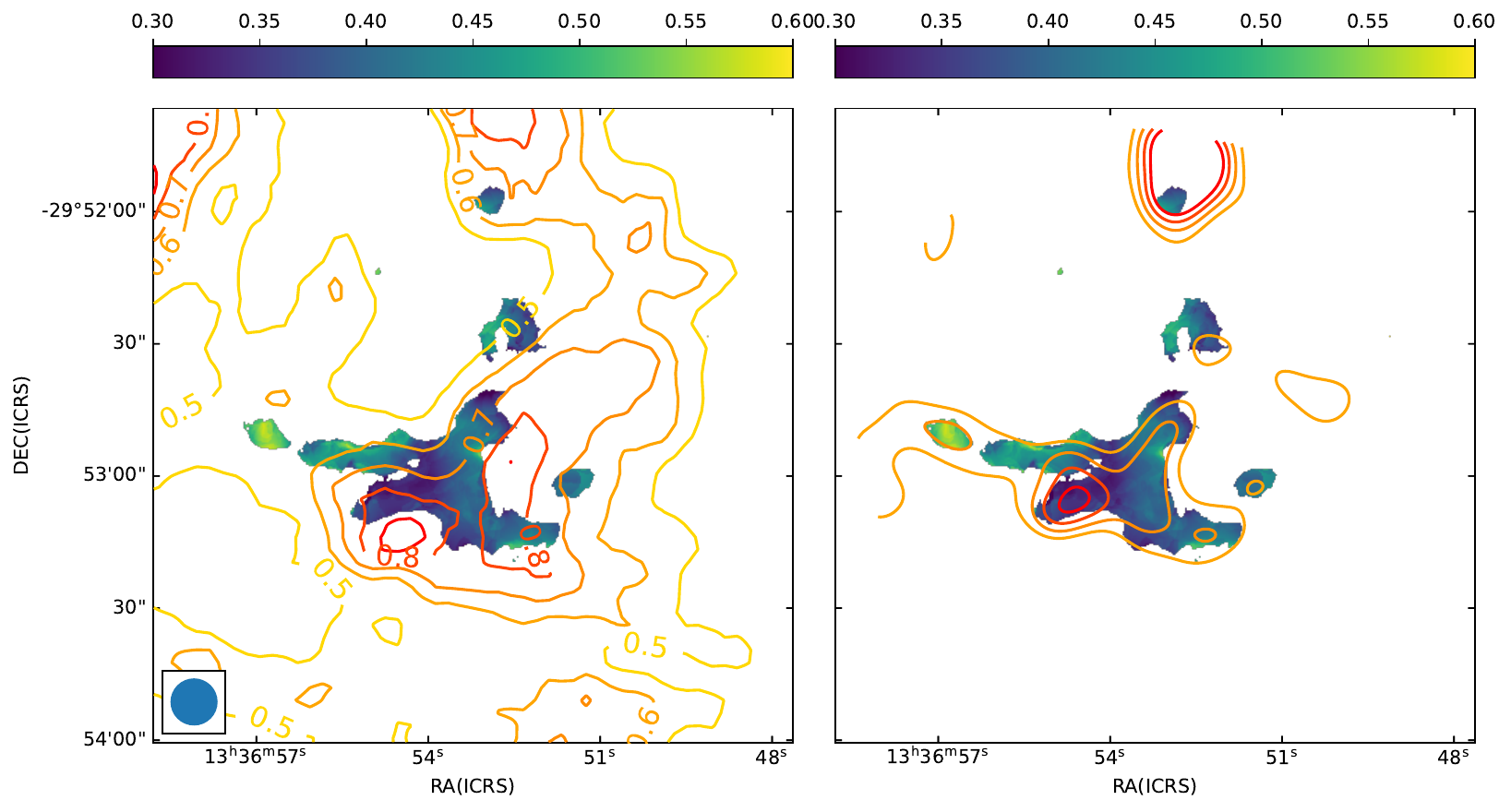}
    \caption{Color image in both panels: HNC/HCN intensity ratio map convolved to 10.7\arcsec\ equivalent to the resolution of the 160$\mu$m Herschel map.  Contours: (Left) Herschel 70/160 $\mu$m ratio. The ratio is taken in the unit of MJy sr$^{-1}$. The contour levels start at 0.5, and are drawn by an increment of 0.1. (Right) The 3-mm continuum convolved to 10.7\arcsec. The contour levels are 0.3, 0.6, 0.9, 1.2 mJy beam$^{-1}$.}
    \label{fig:FIR_ratio}
\end{figure*}

The FIR ratio can be a probe of dust temperature, and an increase in the dust temperature can be caused by UV photons from a star-forming activity. We obtained the FIR data from Herschel at $70\,\mu$m and $160\,\mu$m \citep{2012MNRAS.419.1833B,vngs}, which correspond to dust emission peaks at $\sim 40$ and 20\,K, respectively. Since currently available far-IR observations only have a coarser resolution than the ALMA data, we convolved our ALMA data into the beam size of the 160-$\mu$m Herschel data using the CASA task \texttt{imsmooth}. Since  the Herschel 70-$\mu$m data also has a smaller beam ($5.6\arcsec$) compared with the 160-$\mu$m data, we also convolved the 70-$\mu$m data to the beam size of the 160-$\mu$m data, 10.7$\arcsec$. Figure \ref{fig:FIR_ratio} (left) shows the HNC/HCN ratio at a $10.7\arcsec$ resolution superposed with the ratio of 70$\mu$m over $160\mu$m flux. This HNC/HCN ratio map was created by convolving the HCN and HNC images individually to $10.7\arcsec$, and taking the ratio. The Herschel data have already been presented by \citet{2012MNRAS.421.2917F}. The HNC/HCN ratio tends to be high when the 70$\mu$m/$160\mu$m ratio is lower. This is an expected result if the UV field both heats the dust and decreases the HNC/HCN ratio. Moreover, the HNC/HCN ratio can be low even if the 70$\mu$m/$160\mu$m ratio is moderately low if it is near the high 70$\mu$m/$160\mu$m ratio region. We can interpret this trend as a global (a few 100\,pc scale) effect of the UV field on the HNC/HCN ratio.

Since the FIR data are only available in a lower resolution, it is helpful to compare the HNC/HCN ratio with the star formation rate with the same resolution as the FIR data. Figure \ref{fig:FIR_ratio} (right) is similar to Figure \ref{fig:FIR_ratio} (left), but the contours are the 3-mm continuum convolved to the same resolution as the Herschel beam. The large-scale variation of the 3-mm continuum appears somewhat similar to that of the 70$\mu$m/$160\mu$m ratio, although the 3-mm continuum is more concentrated around regions detected with molecules. The region with the peak of the 70$\mu$m/$160\mu$m ratio has a low HNC/HCN ratio, which may indicate that the global star formation rate impacts the HNC/HCN ratio. At the same time, the HNC/HCN ratio peak at the eastern part of the map has a moderately high star formation rate inferred from the 3mm continuum. There does not seem to be a consistent trend between the star formation rate and the HCN/HCN ratio even globally.

\section{Discussion}\label{sec:disc}
\subsection{Factors to cause variations in the HNC/HCN abundance ratio}\label{sec:factors}
Before discussing the implications of our results, let us summarize the factors that may affect the abundances of HCN and HNC, and the ratio between them.
\subsubsection{Temperature}
It was first suggested that the temperature dependence of the HNC/HCN ratio is due to some reactions to destroy HNC with barriers:
\begin{equation}\label{eq:o_reac}
    \mathrm{HNC + O \longrightarrow NH + CO}
\end{equation}
 and 
 \begin{equation}\label{eq:h_reac}
     \mathrm{HNC + H \longrightarrow HCN + H}.
 \end{equation}
These reactions with barriers become very efficient above certain temperatures because their reaction rates are proportional to exp(-$\gamma/T$), where $\gamma$ is the reaction barrier and $T$ is the gas kinetic temperature. There has been some debate about the values of these barriers. Quantum mechanical calculations show that the barrier of Reaction \ref{eq:o_reac} is 1100\,K \citep{lin1992implications}, while that of Reaction \ref{eq:h_reac} is 2000\,K or 1200\,K \citep{1996A&A...314..688T,2014ApJ...787...74G}. The temperature dependence of HNC/HCN seen in observations requires much lower barriers. \citet{2020A&A...635A...4H} concluded that the barriers of 20\,K and 200\,K provide the best fit to the observed trends. This discrepancy between the calculated reaction rates and observed temperature dependence suggests that we may be missing important reactions in the network, or that the calculated rates may not have enough accuracy, or a physical parameter or process other than the temperature is responsible for the observed temperature dependence of HNC/HCN. 

\subsubsection{FUV radiation}
Recently, \citet{2023A&A...679A...4S} suggested from their observations in the Orion B region that the interstellar radiation field may be the biggest contributing factor to the HNC/HCN ratio rather than the temperature. This is because the FUV radiation promotes both the formation and excitation of HCN. There are additional formation reactions of HCN at $A_{\rm V}<4$ mag.: \({\rm C + HNC \longrightarrow HCN + C}\) and \({\rm N + HCO \longrightarrow HCN + O}\). The higher ionization fraction in PDRs also makes the electron excitation more efficient. The UV radiation should of course increase the temperature, and some relationship between the HNC/HCN ratio and the temperature is expected without the need for low reaction barriers.  

\subsubsection{Cosmic rays}
\citet{2022ApJ...939..119B} studied the dependence of HNC/HCN on the cosmic-ray ionization rate to analyze the chemistry in the starburst galaxy NGC~253. Because the main formation pathway of HCN and HNC with high cosmic-ray ionization rates is
\begin{equation}
    \mathrm{HCNH^+ + e^-} 
    \longrightarrow 
    \begin{cases}
        \mathrm{HCN + H} \\
        \mathrm{HNC + H  }\\
        \mathrm{CN + H + H }
    \end{cases},
\end{equation}
 this reaction maintains the HCN/HNC abundance close to unity. The branching ratio of the electron recombination reaction of HCNH$^+$ is estimated to be roughly 1/3 for each product HCN, HNC, and CN according to the KIDA database\footnote{https://kida.astrochem-tools.org/}.

\subsection{Measured quantities and physical parameters}
In the previous sections, we compared the HNC/HCN ratios with the 3-mm continuum and the 70$\mu$m/160$\mu$m ratios. Here we connect these measurable quantities to the physical parameters that could affect the chemistry, namely, the temperature, UV field, and cosmic-ray ionization rates. 

{\bf The 3-mm continuum emission}

As mentioned above, the 3-mm continuum is considered a tracer of SFR. Although continuum flux could contain synchrotron radiation or dust emission, there is no obvious synchrotron source such as supernova remnants \citep{2022ApJ...929..144L}. Although there may be unidentified diffuse supernova remnants that cause synchrotron radiation, we do not expect it to be more significant than {\bf for} normal star-forming galaxies, where synchrotron radiation is insignificant at the 3-mm wavelength. We already discussed the estimate of the dust contamination in Section \ref{sec:obs}, which tends to be insignificant.

Good spatial correlations with the temperature and SFR have been observed in many regions. Galactic centers have higher temperatures on the GMC scale where the star formation rates are higher \citep[e.g., ][]{2016A&A...586A..50G,2024ApJ...961...18T} according to LVG analyses. An LVG analysis on the data towards the center of M83 using the same method as \citet{2024ApJ...961...18T} shows elevated temperatures ($\gtrsim$ a few hundred K) near regions with intense star formation compared to the region without star formation ($\sim 100\,$K)(Kaneko et al. in preparation).
Such correlation is also seen in the spiral-arm region of the Galaxy, but the heating is only effective on the scale of 1\,pc or so \citep[e.g., ][]{2020A&A...635A...4H,2022MNRAS.510.1106M}. Of course, having a higher number of star-forming regions within a GMC should raise the temperature on average, but it is unclear how much the temperature is raised for a given value of SFR in the extragalactic GMCs where the density structure within the beam is unknown. 

A high star formation rate also produces a high UV radiation field. While we expect some spatial correlation between the star formation rate and the UV field, the correlation between the star formation rate and the UV radiation field may not be tight if the region has high visual extinction that allows the attenuation of the UV photons. 

The correlation between the cosmic-ray ionization rate and the 3-mm continuum should be weak. First, the locations of supernova remnants \citep{2022ApJ...929..144L} do not seem to correlate with the star formation rate on $<100\,$pc scale. Moreover, cosmic rays penetrate high column densities, and variation among GMCs should be small although we do expect $\sim$kpc-scale variations near the galactic center region where supernova remnants are concentrated.

{\bf FIR intensity ratio (70$\mu$m/160$\mu$m)}

We expect some spatial correlation between the UV radiation field and the FIR intensity ratio. The UV radiation heats the dust, which increases the $70\mu$m/$160\mu$m ratio. If the UV field also heats the gas in addition to the dust, the gas kinetic temperature should also be high where the $70\mu$m/$160\mu$m ratio is high. For the same reasons as the 3-mm continuum, we do not expect a good spatial correlation between the FIR intensity ratio and cosmic-ray ionization rates.

\subsection{Possible scenarios}
\begin{figure*}
    \centering
    \includegraphics[width=0.99\linewidth]{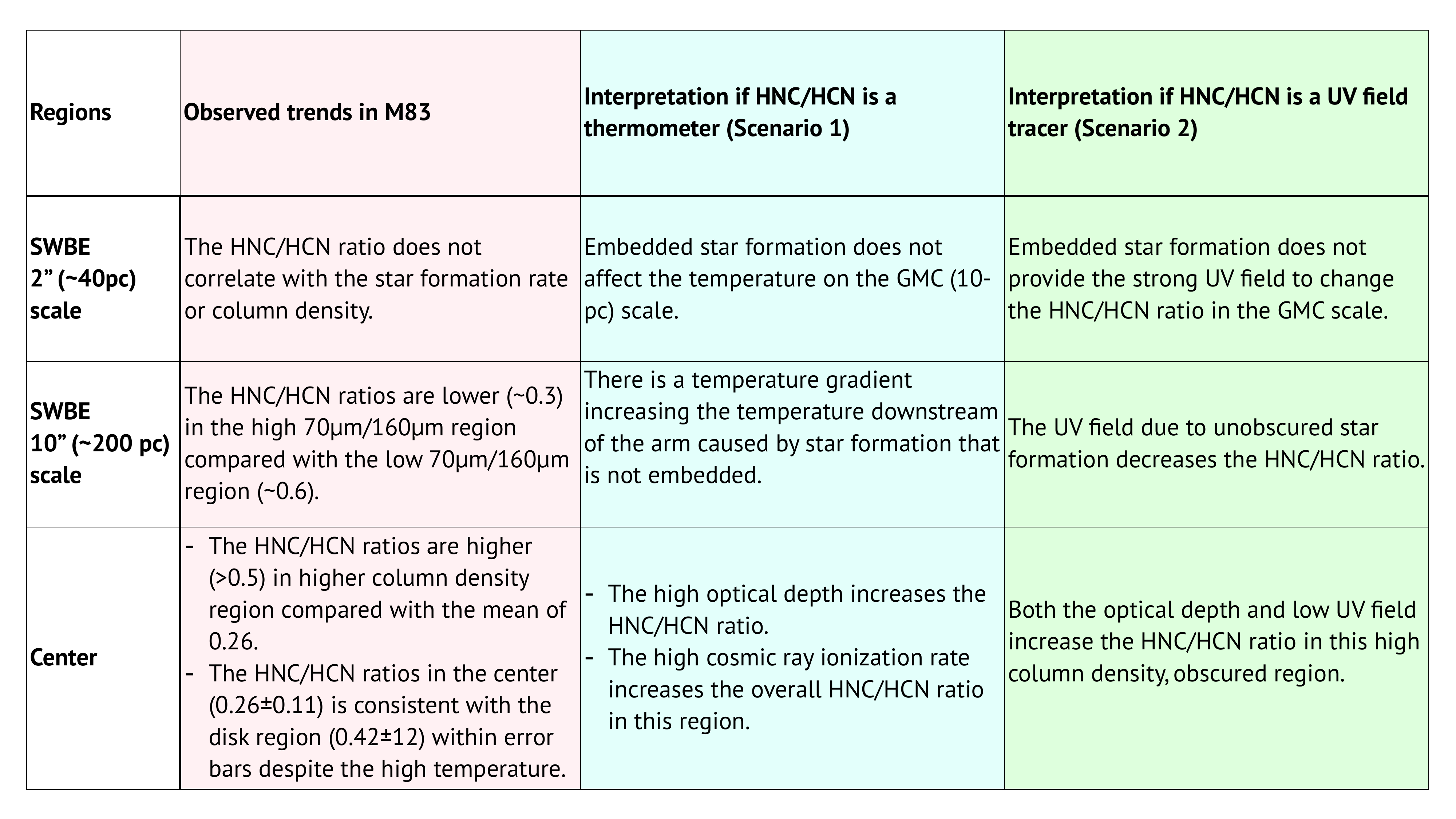}
    \caption{A table describing the interpretation of the observed trends.}
    \label{fig:interpret}
\end{figure*}

We have seen that the HNC/HCN ratio varies with the large-scale variation of 70/160$\mu$m, but not with the local star formation rates in the bar-end region. We also found that the HNC/HCN ratio is largely determined by the optical depth in the galactic center region. Depending on the driving force of the HNC/HCN chemistry, we can propose the following two scenarios. For each scenario, we add the interpretation of observed trends. The summary of this section is also shown in Figure \ref{fig:interpret}.

\begin{enumerate}
    \item {\bf The HNC/HCN ratio changes with the temperature}

    {\bf SWBE $2\arcsec$ (40\,pc) scale}: In this case, the increase in the temperature caused by the high star formation rate is not visible in our HNC/HCN observations. A possible reason is that either the star formation does not increase the temperature or that it occurs on a scale far smaller than our beam size (i.e., parsec scale or less). Because it is hard for star formation not to increase the temperature at all, we speculate that the heated regions are small for the embedded star formation traced with the 3-mm continuum emission.
    
    {\bf SWBE $10\arcsec$ scale ($\sim 200\,$pc)}: There is a gas temperature increase on a large scale as the gas flows through the bar end region, which lowers the HNC/HCN ratio downstream. The heating source is unknown, but possible candidates are photoelectric heating due to UV photons, shock heating, and cosmic-ray heating. If the heating source is UV photons, it is difficult to distinguish whether the driving force of the chemistry is temperature or UV photons. \citet{2020ApJ...890L..10K} found that the CO(2--1)/(1--0) ratio is higher downstream around spiral arms, and proposed that this higher ratio is caused by UV or cosmic-ray heating due to star formation that took place when the gas passed through the spiral arms. This star formation causing the temperature increase must be unobscured instead of the embedded one. Cosmic rays are unlikely to be the main driver of chemistry in regions with high CO(2--1)/(1--0) ratios because the cosmic rays increase the HNC/HCN ratio.

    {\bf The central region}: The center of M83 has a temperature of $>100\,$K according to the LVG analysis by Kaneko et al. (in preparation). Most of the gas is supposed to have HNC/HCN$<0.1$ if this ratio follows the relationship by \citet{2020A&A...635A...4H}. We can interpret the high HNC/HCN ratio with the high cosmic-ray ionization {\bf rate}. There is still an effect of the optical depth that alters the HNC/HCN ratio locally. 
    
    \item {\bf The HNC/HCN ratio changes with the UV radiation field}

    {\bf SWBE $2\arcsec$ scale}: If the HNC/HCN ratio is affected by the UV radiation field, its effect is limited in highly obscured regions. Star-forming regions traced with the 3mm continuum may not alter the HNC/HCN ratio because these star-forming regions are embedded, providing little dependence of the HNC/HCN on the star formation rate in the bar-end region on the small scale.
    
    {\bf SWBE $10\arcsec$ scale}: The global dependence of the HNC/HCN ratio on the far-infrared ratio is caused by the UV radiation field because the UV radiation also heats the dust. 
    
    {\bf The central region}: In the galactic center region, the HNC/HCN ratio is not very low despite the high star formation rate and high temperature because the density is also high, and the UV radiation field can be attenuated. The high HNC/HCN ratio in the obscured region is caused by both the lack of UV field and the optical depth.  
\end{enumerate}

It is difficult to determine which scenario is correct from the observations in the SWBE region alone, especially if the UV photons are the main heating source. To decide which scenario is more likely, it is important to study regions where the heating source is not UV photons, i.e., regions that are obscured from external UV photons. The spatial distribution of the HNC/HCN ratio in the galactic center region can give us a clue, since it generally has high column densities judging from the fact that molecular emission and dust continuum emission are both brighter than the bar-end region by about an order of magnitude. For the HNC/HCN ratio to be high enough in the high-temperature region of the M83 galactic center, cosmic-ray ionization needs to be high according to Scenario 1, where the HNC/HCN ratio is mainly determined by the temperature. While the cosmic-ray ionization rate is expected to be high due to the high supernova rate \citep{2022ApJ...929..144L}, it is unlikely to be the main driver of the chemistry. If cosmic rays increase the HNC/HCN ratio, they should have a stronger impact in the low-density region, because the chemistry scales as $\zeta/n$, where $\zeta$ is the cosmic-ray ionization rate and $n$ is the density. This contradicts the observed trend in M83. The HNC/HCN ratios are also high in high-column-density regions in the central region of NGC~253 \citep{2022ApJ...939..119B}. 

Even in Scenario 1, one could also argue that the HNC/HCN ratio is high in the central region simply because of the optical depth despite the high temperature. However, the HNC/HCN ratio is moderately high ($\sim 0.3$) even where a line is broad and likely optically thin around the optical nucleus (near the position 1 in Figure \ref{fig:mom0_cen}). The temperature corresponding to $I$(HNC)/$I$(HCN)=0.3 is about 30\,K according to the \citet{2020A&A...635A...4H} relation, and is too low for a galactic center. Although it appears reasonable to exclude this case, deep maps of H$^{13}$CN and an estimate of $^{12}$C/$^{13}$C are necessary for confirmation so that the optical depth in this whole region can be derived.

It may sound unreasonable that the HNC/HCN ratio is moderately high in the galactic center region despite the high UV field expected from the high star formation rate if Scenario 2 is true. However, the HNC/HCN ratio is low ($\lesssim 0.2$) in low column density regions, where the UV photons can travel without too much extinction. Near-infrared observations towards the galactic center region reveal higher visual extinction \citep[$A_\mathrm{V}\gtrsim 14$;][]{1987ApJ...313..644T}. Molecular regions likely have even higher visual extinction, which would increase the HNC/HCN ratio if the HNC/HCN ratio decreases with the UV field. For these reasons, we believe that Scenario 2 is more plausible.

\section{Summary}\label{sec:summary}
We present 40-pc or 50-pc resolution dense gas tracers (HCN, HNC, and \hcop) in the southwestern bar-end and central regions of M83 obtained by ALMA. We focus on the isomeric ratio of HNC and HCN, which has been previously proposed as a temperature tracer or more recently as a UV field tracer. Here is the summary of our findings.
\begin{itemize}
    \item The southwestern bar end and the center of M83 lie on the typical relationship between the dense gas surface density ($\Sigma_\mathrm{dense}$) and the star formation surface density ($\Sigma_\mathrm{SFR}$) similar to other star-forming galaxies (e.g., M51, NGC~1068) found in the literature.
    \item The HNC/HCN ratio has no obvious correlation with the dense gas surface density, star formation rate, and star formation efficiency in the southwestern bar-end region on the GMC scale ($\sim 40\,$pc).
    \item The HNC/HCN ratio slightly increases with the dense gas surface density, star formation rate, and star formation efficiency in the central region. This is the opposite of what is expected if we assume that the HNC/HCN ratio decreases with increasing temperature and that the high star-formation region raises the kinetic temperature.
    \item Spectra of the central region, especially regions with high HNC/HCN ratios, show a double-peak or flat-top shape for HCN while they are closer to Gaussian for HNC. This trend suggests that these regions are optically thick in HCN. High optical thickness explains the high HNC/HCN ratios in high dense gas surface density, high star-formation-rate regions.
    \item The HNC/HCN ratio tends to be lower near the region where the ratio of $70\mu$m/$160\mu$m luminosity is high on a large scale of $\sim 200\,$pc in the southwestern bar-end region. The low HNC/HCN ratios tend to appear downstream across the bar-end. This result suggests that star formation on the bar/arm either raises the temperature or increases the UV radiation field not on the GMC scale, but on a larger scale.
    \item  The above results suggest that the HNC/HCN ratio is likely determined by the UV radiation field. It has a dependence on the large scale because UV photons travel farther in low column density regions in the bar-end region. It has little dependence on star formation rate because star formation in the bar-end region detected by the 3-mm continuum is mostly embedded. The column density of the galactic center region is very high, and the HNC/HCN ratio is high despite the high star formation rate in this region. The cosmic-ray ionization rate is likely high, but is unlikely the driver of the HNC/HCN chemistry. 

\end{itemize}

Although our conclusion is that the HNC/HCN ratio is a UV field tracer, it is necessary to have more careful analysis in regions where the main heating source is not UV photons and where other temperature tracers are available to confidently distinguish whether it is a temperature tracer or a UV tracer. It is also important to note that the HNC/HCN ratio may have a temperature dependence when the temperature is close enough to the reaction barrier of conversion reactions from HNC to HCN (e.g., $\gtrsim$ a few 100\,K in the shocked region).
\begin{acknowledgments}
We thank the anonymous referee for the helpful comments. This paper makes use of the following ALMA data: \\
ADS/JAO.ALMA\#2016.1.00164.S\\
ADS/JAO.ALMA\#2015.1.01177.S\\
ADS/JAO.ALMA\#2021.1.01195.S. \\
ALMA is a partnership of ESO (representing its member states), NSF (USA) and NINS (Japan), together with NRC (Canada), MOST and ASIAA (Taiwan), and KASI (Republic of Korea), in cooperation with the Republic of Chile. The Joint ALMA Observatory is operated by ESO, AUI/NRAO and NAOJ. N.H. acknowledges support from JSPS KAKENHI grant No. JP21K03634. Y.N. gratefully acknowledges support from JSPS KAKENHI grant Numbers JP23K13140 and JP23K20035. K.S. thanks support by the grants NSTC 111-2124-M-001-005- and NSTC 112-2124-M-001-014-.
\end{acknowledgments}

%

\vspace{5mm}
\facilities{ALMA, Herschel}


\software{astropy \citep{2013A&A...558A..33A,2018AJ....156..123A}, CASA \citep{2007ASPC..376..127M, 2022PASP..134k4501C} 
          }





\bibliography{sample631}{}
\bibliographystyle{aasjournal}



\end{document}